\documentclass[%
reprint,
 amsmath,amssymb,
 aps,
]{revtex4-2}

\usepackage{graphicx}
\usepackage{dcolumn}
\usepackage{bm}
\usepackage{placeins}

\begin{document}

\title{Optimized strategies for the quantum-state preparation of single trapped nitrogen molecular ions}


\author{Aleksandr Shlykov}
\author{Mikolaj Roguski}
\author{Stefan Willitsch}
\email{stefan.willitsch@unibas.ch}

\affiliation{%
Department of Chemistry, University of Basel, 4056 Basel, Switzerland}%

\date{\today}

\begin{abstract}
This work examines optimized strategies for the preparation of single molecular ions in well-defined rotational quantum states in an ion trap with the example of the molecular nitrogen ion N$_2^+$. It advances a two-step approach consisting of an initial threshold-photoionization stage which produces molecular ions with a high probability in the target state, followed by a measurement-based state purification of the sample. For this purpose, a resonance-enhanced threshold photoionization scheme for producing N$_2^+$ in its rovibrational ground state proposed by Gardner et al. [Sci.~Rep.~\textbf{9},~506~(2019)] was characterized. The molecular state was measured using a recently developed quantum-non-demolition state-detection method finding a total fidelity of 38$\pm$7\% for producing ground-state N$_2^+$ under the present experimental conditions. By discarding ions from the trap not found to be in the target state, essentially state-pure samples of single N$_2^+$ ions can be generated for subsequent state-specific experiments. 

\end{abstract}


\maketitle

\section{\label{sec:level1}Introduction}
In recent years, molecular-state detection using quantum-logic techniques opened up new horizons for the coherent manipulation of single molecular ions in traps~\cite{wolf16a, chou17a, sinhal20a}. Compared to previously employed destructive methods~\cite{koelemeij07a, staanum10a, tong10a, germann14a}, these approaches provide vastly higher experimental duty cycles and, therefore, higher measurement statistics and levels of precision. They unfold novel perspectives for utilizing precision-spectroscopic experiments to probe fundamental physics~\cite{demille17a, safronova18a}, for precisely determining values of fundamental constants~\cite{biesheuvel16a, alighanbari18a} and for testing their possible temporal variation~\cite{schiller05a, kajita14a}, for developing new frequency standards based on molecular rovibrational transitions~\cite{schiller14a}, for employing molecular ions as qubits~\cite{najafian20b} and for observing and controlling chemical reactions of single particles on the quantum level~\cite{najafian20a}.

A necessary step in any coherent single-molecule experiment is the initialization of the molecule in a well-defined quantum state, often its electronic, vibrational, and rotational (rovibrational) ground state. Over the past decade, various approaches have been implemented towards that purpose. Polar molecular ions can be optically pumped to the rovibrational ground state using dipole-allowed spectroscopic transitions~\cite{staanum10a, schneider10a, lien14a}. However, polar species are also exposed to a constant redistribution of level populations by ambient black body radiation and, therefore, require a cryogenic environment to preserve quantum states. Another potential approach is cryogenic-buffer-gas cooling which also requires a cryogenic setup~\cite{hansen14a}. Moreover, this technique is not entirely state selective because it leaves the ions in a (low-temperature) thermal distribution of level populations. In this case, the state purity of the sample can be increased by selectively discarding ions which are not in the target state~\cite{hansen14a, hauser15a}.

Resonance-enhanced multi-photon ionization (REMPI) is a widely applicable and highly selective method for the generation of molecular ions~\cite{tong10a, opitz90a, mackenzie95a, malow01a, penno04a, schmidt20a}. Combined with threshold-photoionization techniques, i.e., using photon energies just above the ionization threshold of the neutral molecule, this approach, in principle, provides high state-preparation fidelities for molecular ions~\cite{mackenzie95a, tong10a, zhang23a}. However, it is sensitive to external electric fields which shift the ionization thresholds of different rotational states~\cite{zhang23a, merkt97a, blackburn20a}. 

In the present paper, we focus on optimized state-preparation strategies for single molecular nitrogen ions in radiofrequency (RF) ion traps. N$_2^+$ has been proposed as an attractive system for precision molecular spectroscopy as it features small systematic shifts, transitions with low sensitivity to magnetic fields and, as an apolar diatomic ion, its rovibrational levels in the electronic ground state possess a natural immunity to black-body radiation~\cite{najafian20b, kajita15a}. We have previously achieved the preparation of N$_2^+$ in specific rotational states in an ion trap by employing a $[2+1']$ REMPI scheme~\cite{tong10a, tong11a}. 

\begin{figure*}[ht!]
\includegraphics[width=1\textwidth]{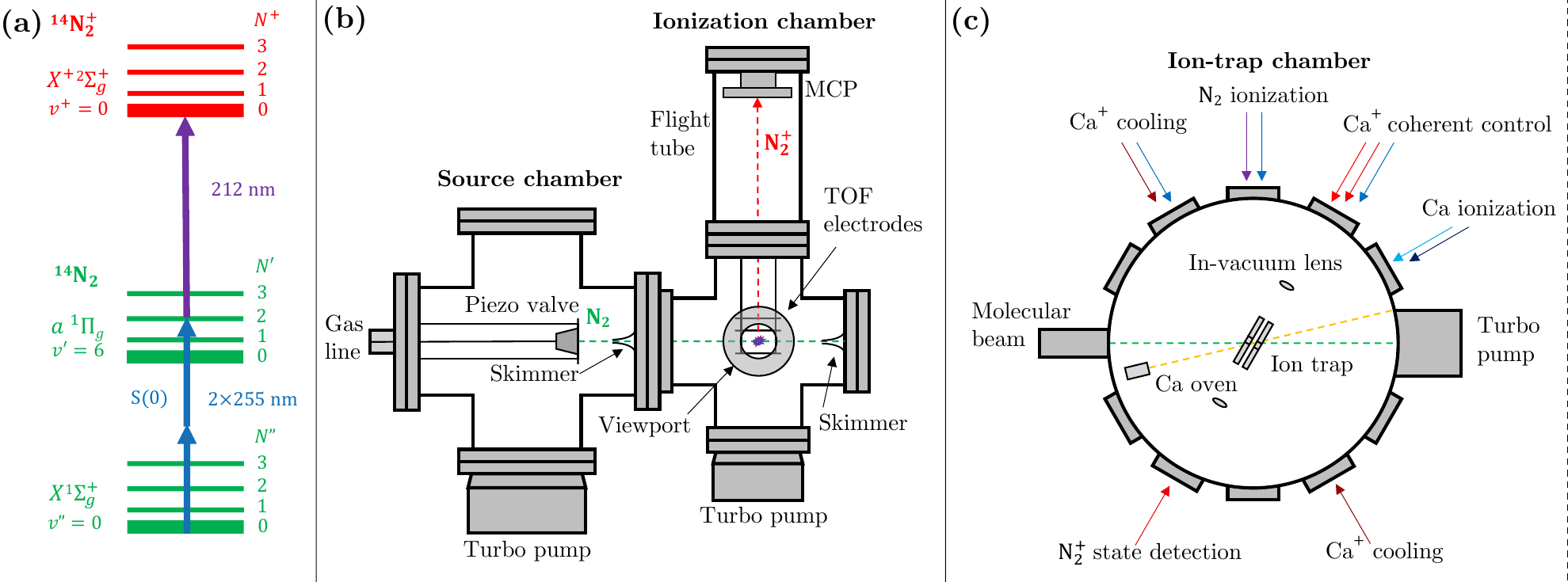}
\caption{a) Energy-level diagram of the $[2+1']$ REMPI scheme employed for the rovibrational-ground-state preparation of single N$_2^+$ ions~\cite{gardner19a}. b) Schematic of the setup used for the REMPI TOF-MS measurements. c) Ion-trap experiment employed in the present study. Colored arrows represent different laser beams used for ionization, cooling, and the coherent control of atomic and molecular ions. Dashed lines indicate atomic (yellow) and molecular (green) beams. See text for details.}
text for details.
\label{fig:expt}
\end{figure*}

However, the REMPI scheme used in Reference~\cite{mackenzie95a} suffers from spurious [2+1] one-color photoionization which, if not carefully suppressed, diminishes the state selectivity of the experiment. To mitigate this problem, an alternative $[2+1']$ REMPI scheme via the $a^1\Pi_g (v'=6)$ intermediate state of N$_2$ was put forward by Gardner et al.~\cite{gardner19a} (Figure~\ref{fig:expt}a). In this approach, two photons at 255~nm are required for the excitation to the intermediate state and another photon at $212$~nm for ionization. As the energy of the individual photons in the excitation step is smaller than the one for ionization, parasitic [2+1] ionization is suppressed.

In the present work, we evaluated this alternative REMPI scheme as a basis for the preparation of single N$_2^+$ ions in their rovibrational ground state within a radiofrequency ion trap. Using a recently developed, highly sensitive quantum-non-demolition (QND) state readout scheme~\cite{sinhal20a}, we demonstrate a total state-preparation fidelity of $38\pm7$~\% which is limited by RF-field-induced time-varying shifts of the ionization thresholds of N$_2$ and secondary ionization processes. Because of the non-destructive nature of the detection scheme, the state of the ion is preserved and ions not found in the target state can be discarded from the trap. Only ions in the desired state are kept which can be utilized for subsequent state-specific experiments. We advocate the presently adopted approach, i.e., threshold photoionization combined with subsequent QND state detection and post-selection of the ions, as a general technique for producing single trapped molecular ions in specific quantum states with essentially unit fidelity.


\section{\label{sec:Methods}Methods}
The present work comprised two sets of experiments. First, REMPI and photoionization spectra of N$_2$ via the \mbox{$a^1\Pi_g(v'=6)$} $\leftarrow$ \mbox{$X^1\Sigma_g^+(v''=0)$} transition~\cite{gardner19a} were recorded in a dedicated molecular-beam setup as a prerequisite for the implementation of the photoionization scheme in an ion-trap apparatus. Second, photoionization was subsequently carried out inside the trap and combined with a recently developed QND state detection method~\cite{sinhal20a,meir19a} for both evaluating the state-selectivity of the photoionization and simultaneously performing projective state preparation of the ions.


\subsection{\label{subsec:REMPI} REMPI and photoionization spectroscopy of jet-cooled N$_2$}
The $[2+1']$ REMPI scheme (Figure~\ref{fig:expt}a) was investigated using a molecular beam machine coupled to a Wiley-McLaren~\cite{wiley55a} time-of-flight mass spectrometer (Figure~\ref{fig:expt}b). 

A beam of internally cold nitrogen molecules was produced by supersonic expansion of N$_2$ gas from a pulsed piezo valve (Amsterdam Piezo Valve, MassSpecpecD BV) at 2 bar backing pressure and a repetition rate of 10~Hz. The beam passed a skimmer with 1~mm diameter (Model 2, BeamDynamics) before entering the ionization chamber at a base pressure of $3\times$10$^{-9}$~mbar. 

For the REMPI experiments, the output of two dye lasers (NarrowScan, Radiant Dyes) pumped by the third harmonic of an Nd:YAG laser (Spitlight 1500, Innolas) was used. One dye laser producing light at 255~nm after frequency doubling was employed to excite N$_2$ to the intermediate state, and the other frequency-doubled laser producing light at 212~nm was used for the ionization step. Laser wavelengths were measured using a wavemeter with an internal calibration source (WS-6, High Finesse). The powers of the excitation and ionization lasers were maintained at 1~mJ per pulse and 0.75~mJ per pulse, respectively, measured before entering the chamber. The two lasers were focused into the molecular beam and ions were extracted into the TOF tube 1~$\mu$s after ionization to be detected by a multi-channel plate (MCP) detector (APD-APTOF, Photonis).

Spectra were recorded by monitoring the integrated nitrogen ion signal as a function of the relevant laser wavenumber. For measuring spectra of the \mbox{$a^1\Pi_g (v'=6)$}$\leftarrow$\mbox{$X^1\Sigma_g^+(v''=0)$} two-photon transition in N$_2$, the frequency of the ionization laser was fixed at 47175~cm$^{-1}$ and the excitation laser frequency was scanned within a 40~cm$^{-1}$ interval. To record photoionization spectra, the frequency of the excitation laser was fixed to the S(0), S(1), and Q(1) rotational components of \mbox{$a^1\Pi_g (v'=6)$}$\leftarrow$\mbox{$X^1\Sigma_g^+(v''=0)$} transition, and the frequency of the ionization laser was scanned.

\subsection{\label{subsec:QND}Quantum-logic-spectroscopic characterization of the photoionization products and projective state preparation of single N$_2^+$ ions in the ion trap}

The setup for quantum-logic spectroscopy of single trapped molecular ions consisted of a linear RF ion trap in an ultrahigh-vacuum chamber (base pressure $2\times10^{-11}$~mbar), shown schematically in Figure~\ref{fig:expt}c, coupled to a molecular-beam machine through a skimmer with an orifice diameter of 0.5~mm (Model 2, BeamDynamics). The inscribed radius of the RF electrodes of the trap amounted to $r=1.75$~mm and the distance between the endcap electrodes was $d=5$~mm. 
Coulomb crystals of Doppler-laser-cooled Ca$^+$ ions were loaded into the trap from a skimmed atomic beam of neutral Ca atoms by a resonant photoionization scheme~\cite{lucas04a}, followed by sympathetically cooling a single N$_2^+$ ion produced by the threshold-photoionization scheme discussed above. During Ca$^+$ loading and N$_2$ ionization, the RF ion trap was operated at the frequency $\Omega_{RF}=2\pi\times16.4$~MHz, an RF amplitude $V_{RF}\approx516$~V and a potential of $V_{DC}=15$~V applied to the endcap electrodes. The axial frequency of an N$_2^+$ ion corresponding to these parameters was $\omega=243$~kHz and the Mathieu q-parameter \cite{willitsch12a} was $q=0.11$. After reducing the crystal to a N$_2^+$-Ca$^+$ two-ion string~\cite{meir19a}, the ions were cooled to the motional ground state of their axial center-of-mass motion in the trap by repeatedly driving a red motional sideband of the \mbox{(4s)~$^2$S$_{1/2}$} $\rightarrow$\mbox{(3d)~$^2$D$_{5/2}$} clock transition at 729~nm in Ca$^+$~\cite{meir19a}. 

\begin{figure}[t]
\includegraphics[width=0.5\textwidth]{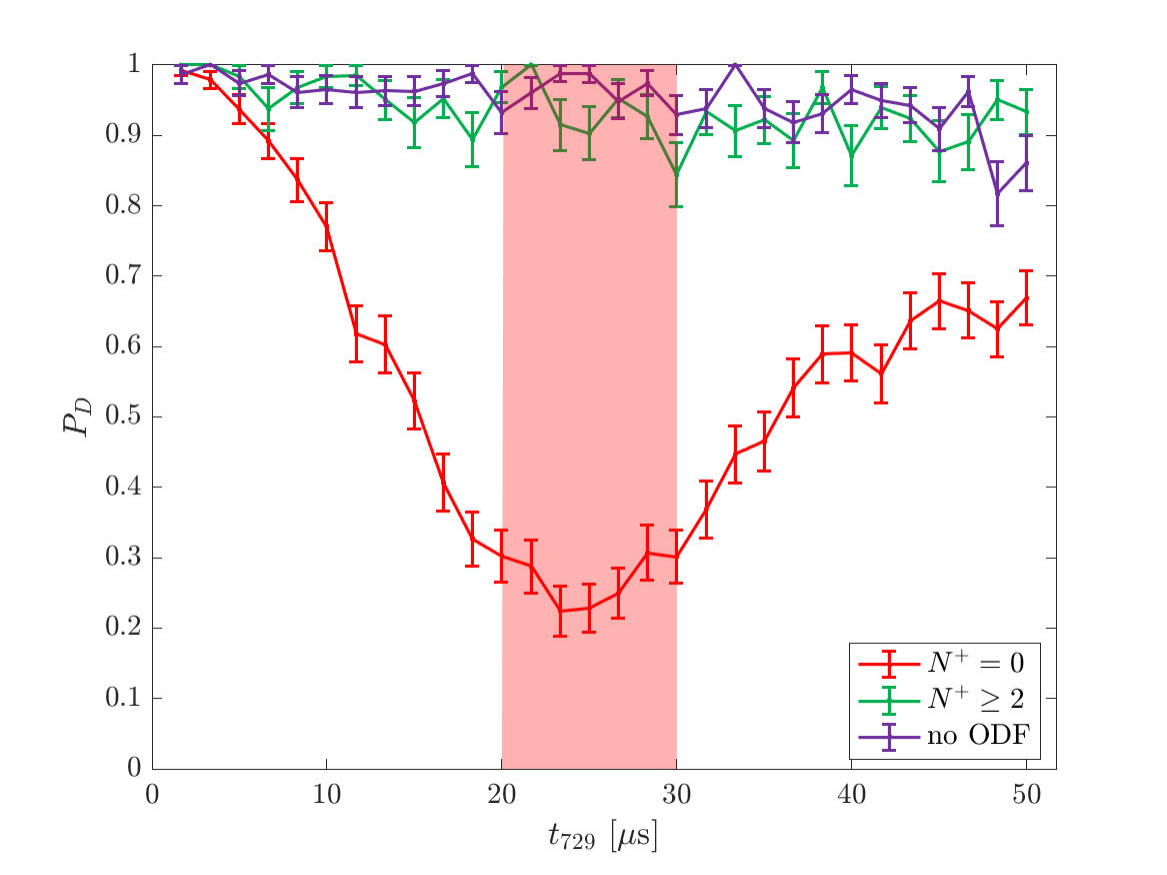}
\caption{Population of the excited state $P_D$ during Rabi flops on the \mbox{(3d)~$^2$D$_{5/2}, m=-5/2$} $\rightarrow$ \mbox{(4s)~$^2$S$_{1/2}, m=-1/2$} blue motional sideband of the Ca$^+$ clock transition following coherent motional excitation of a Ca$^+$-N$_2^+$ two-ion string by a state-dependent optical dipole force.  The red and green traces show experiments with N$_2^+$ in its rovibrational ground state ($N^+$=0) and higher-lying rotational states ($N^+\geq$2), respectively. The purple trace represents the signal obtained without the application of an ODF. The red-shaded area indicates the interval of 729 nm laser pulse lengths for which maximum stated detection contrast was achieved. Error bars represent the standard deviation of 175, 75, and 150 measurements for the red, green, and purple traces. See text for further details.}
\label{fig:state_detection}
\end{figure}

The quantum-non-demolition state-detection scheme employed to verify the rovibrational state of the N$_2^+$ ion and thus the state-preparation fidelity of the photoionization has been described in detail previously~\cite{sinhal20a}. Briefly, an optical dipole force (ODF) was generated by a one-dimensional optical lattice of two counter-propagating laser beams around 787~nm detuned from each other by the frequency of the in-phase axial motional mode of the N$_2^+$-Ca$^+$ string in the trap (680.5~kHz in the present experiments). Under these conditions, a strong ODF was generated for N$_2^+$ in the rovibrational ground state exciting the axial center-of-mass motion of the ions. During the application of the optical lattice, the Ca$^+$ ion was shelved in the \mbox{(3d)~$^2$D$_{5/2}, m=-5/2$} state to prevent spurious motional excitation from an ODF generated on the atom~\cite{najafian20a}. The coherent motional excitation was detected by Rabi thermometry on the \mbox{(3d)~$^2$D$_{5/2}, m=-5/2$} $\rightarrow$ \mbox{(4s)~$^2$S$_{1/2}, m=-1/2$} blue motional sideband of the Ca$^+$ 729~nm clock transition, see the red trace in Figure~\ref{fig:state_detection} as a representative example. For all other rovibrational states of N$_2^+$, the optical lattice was too far detuned from any spectroscopic transition to generate a strong ODF, thus leaving the two-ion string in the motional ground state from which no Rabi flops on the sideband transition could be observed (green trace in Figure~\ref{fig:state_detection}). The fidelity of the scheme for detecting the ion in the rotational ground state was demonstrated to be $>99$\%~\cite{sinhal20a}.

\section{\label{sec:Results}Results}

\begin{figure}[t]
\includegraphics[width=0.5\textwidth]{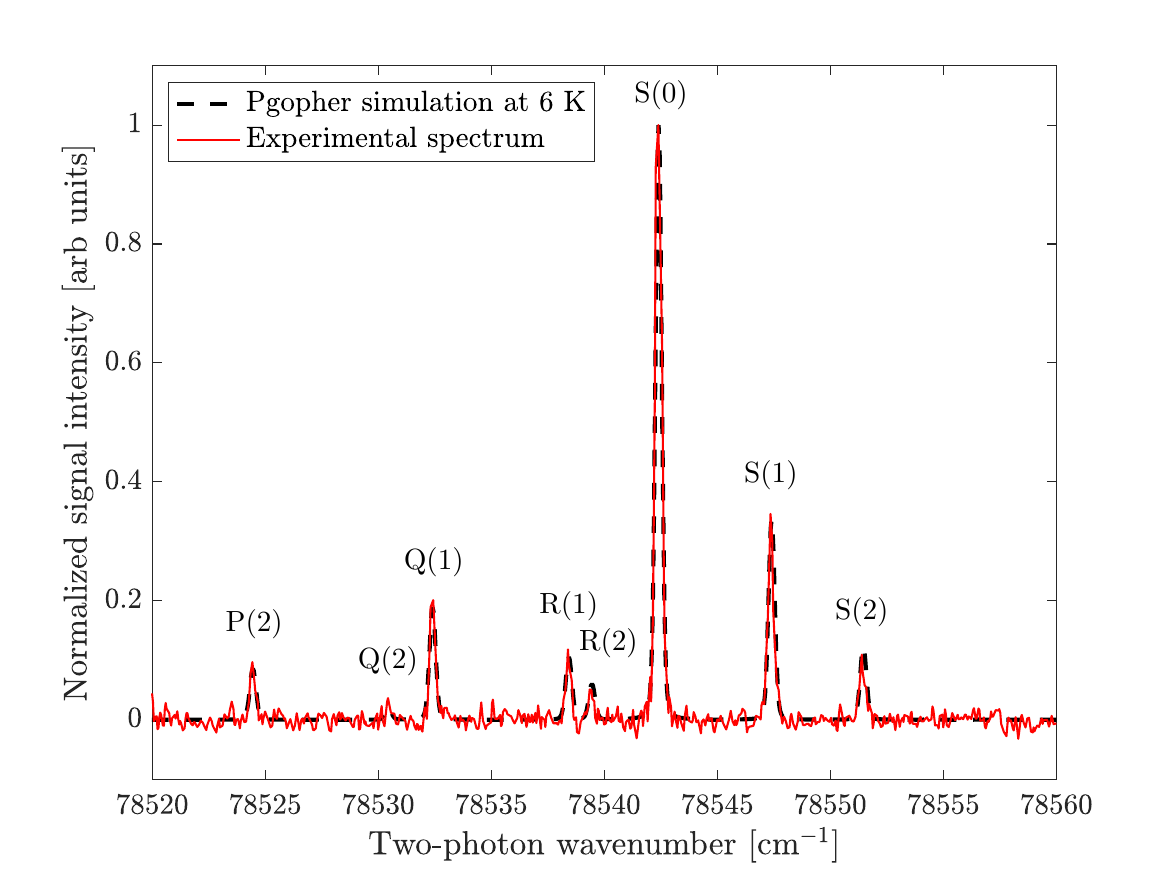}
\caption{$[2+1']$ REMPI spectrum of the \mbox{$a^1\Pi_g(v'=6)$} $\leftarrow$ \mbox{$X^1\Sigma_g^+(v''=0)$} two-photon transition in N$_2$ recorded in a molecular beam. Rotational components of the transition are indicated with standard spectroscopic notation~\cite{herzberg89a}. The black dashed line shows the simulation of the two-photon transition assuming a rotational temperature 6~K and spectroscopic constants from Reference~\cite{huber77a} with the band origin shifted by -0.5~cm$^{-1}$.}
\label{fig:REMPI_spectrum}
\end{figure}

A rotationally resolved spectrum of the \mbox{$a^1\Pi_g(v'=6)$} $\leftarrow$ \mbox{$X^1\Sigma_g^+(v''=0)$} two-photon transition of jet-cooled N$_2$ recorded by scanning the excitation-laser and fixing the ionization-laser wavelength is shown in the red trace in Figure~\ref{fig:REMPI_spectrum}. Eight transitions corresponding to rotational components within the P, Q, R, and S branches were observed in the spectrum. The transition wavenumbers determined from the spectrum agree well with previous results from room-temperature spectroscopy~\cite{gardner19a, vanderslice65a}. The black dashed trace in Figure~\ref{fig:REMPI_spectrum} shows a simulation of the spectrum using PGOPHER~\cite{western17a}. Spectroscopic constants for the simulation were taken from the NIST database~\cite{huber77a} where the band origin had to be shifted by -0.5~cm$^{-1}$ to achieve the agreement with the experimental spectrum shown in the figure. The transition wavenumbers are summarized in Table~\ref{tab:Transition_values}. The best agreement of the simulated with the observed rotational intensities is obtained assuming a rotational temperature of 6~K for N$_2$ in the molecular beam.

\begin{table}[ht]
\caption{\label{tab:Transition_values}%
Wavenumbers of rotational components of the \mbox{$a^1\Pi_g(v=6)\leftarrow X^1\Sigma_g^+(v=0)$} transition in N$_2$ compared to literature and simulated values. See text for details.}
\begin{ruledtabular}
\begin{tabular}{llll}
Transition & Experiment & Literature\footnotemark[1] & Simulation\footnotemark[2]\\
\hline
P(2) & 78524.00 & 78524.33 & 78523.96 \\
Q(2) & 78530.44 & 78530.18 & 78529.96 \\
Q(1) & 78532.44 & 78532.36 & 78531.92 \\
R(1) & 78538.40 & 78538.31 & 78537.92 \\
R(2) & 78539.36 & 78539.26 & 78538.95 \\
S(0) & 78542.40 & 78542.13 & 78541.90 \\
S(1) & 78547.36 & 78547.20 & 78546.91 \\
S(2) & 78551.40 & 78551.23 & 78550.95 \\
\end{tabular}
\end{ruledtabular}
\footnotetext[1]{Reference~\cite{vanderslice65a}}
\footnotetext[2]{PGOPHER simulation \cite{western17a} with constants from \cite{huber77a}}
\end{table}

\begin{figure}[t]
\includegraphics[width=0.5\textwidth]{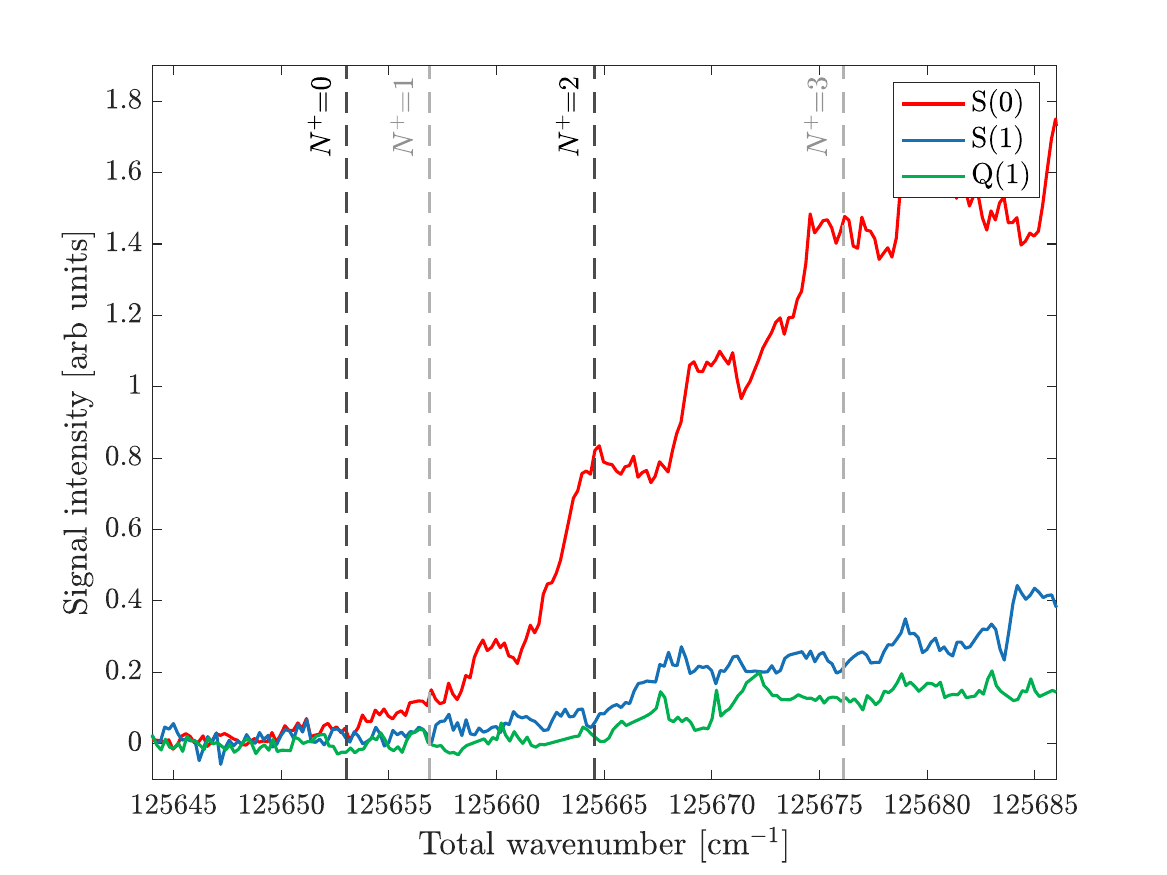}
\caption{Photoionization spectra of N$_2$ recorded via the S(0) (red trace), S(1) (blue trace), and Q(1) (green trace) rotational components of the \mbox{$a^1\Pi_g(v'=6)$} $\leftarrow$ \mbox{$X^1\Sigma_g^+(v''=0)$} band. Approximate ionization thresholds for even and odd rotational states, shifted by 14~cm$^{-1}$ with respect to the literature values to account for stray electric fields in the apparatus, are indicated by black and grey dashed lines, respectively. See text for details.}
\label{fig:Ionisation_threshold_spectrum}
\end{figure}

Figure~\ref{fig:Ionisation_threshold_spectrum} shows photoionization spectra recorded by setting the excitation laser wavenumber to the S(0), S(1), and Q(1) rotational components of the \mbox{$a^1\Pi_g(v'=6)$} $\leftarrow$ \mbox{$X^1\Sigma_g^+(v''=0)$} transition and monitoring the total ion yield as a function of the photoionization-laser wavenumber. The dashed vertical lines indicate approximate ionization thresholds for different ionic rotational states in the present experiment. Because of nuclear-spin-symmetry conservation, ionization from even (odd) intermediate rotational levels only populates even (odd) rotational states in N$_2^+$~\cite{signorell97c}. 
The spacing between the rotational thresholds was calculated using spectroscopic constants for N$_2^+$ from~\cite{huber77a}. The theoretical values were shifted by 14~cm$^{-1}$ compared to the N$_2^+$ ionization energy previously reported in the literature~\cite{huber90a} to account for a lowering of the ionization energies caused by a stray electric field estimated to be $E_{str}\approx5.3$~V/cm in our chamber~\cite{merkt97a}. 

For the generation of N$_2^+$ ions in their rotational ground state in the ion trap, the S(0) component of the \mbox{$a^1\Pi_g(v'=6)$} $\leftarrow$ \mbox{$X^1\Sigma_g^+(v''=0)$} band was excited and the ionization laser wavenumber was set to 47130~cm$^{-1}$, yielding a total ionization wavenumber of 125672.4~cm$^{-1}$ in between the nominal ionization thresholds corresponding to the $N^+=0$ and $N^+=2$ rotational levels of N$_2^+$. The pulse energies of the excitation and ionization lasers were maintained at 1.1~mJ and 0.5~mJ per pulse, respectively, focused to a spot of $\approx 150~\mu$m diameter in the center of the trap. A single N$_2$ molecule from the beam was ionized with a single pair of laser pulses with a probability of 59$\pm$6$~\%$ with these parameters. A contribution of 11$\pm$2$~\%$ from one-color [2+2]-photon ionization~\cite{gardner19a} was determined under these conditions by blocking the ionization-laser beam and monitoring the ion yield. The resonant nature of the ionization process in the trap was verified by changing the frequency of the excitation laser to 78560~cm$^{-1}$, i.e., far detuned from spectroscopic transitions in N$_2$ (see Figure~\ref{fig:REMPI_spectrum}). Under these conditions, the non-resonant ionization probability, attributed to mainly originating from electron impact ionization by spurious electrons ejected from metal surfaces by stray laser light and accelerated by the electric fields in the trap, was measured to be 7$\pm$2$~\%$. These results are summarized in Table~\ref{tab:Resonanse_proove}.

\begin{table}[t]
\caption{\label{tab:Resonanse_proove} Two- and one-color resonant and off-resonant ionization probabilities per laser pulse for the production of single N$_2^+$ ions in the RF ion trap. The last column indicates the number of single laser pulses used in these measurements.}
\begin{ruledtabular}
\begin{tabular} {llll}
Method & Ion. rate p.p. $[\%]$  & Std. error $[\%]$ & Pulses\\\hline
2-color res. ion.& 59 & \(\pm\) 6 & 162 \\
1-color res. ion. & 11 & \(\pm\) 2 & 285 \\
2-color off-res. ion. & 7 & \(\pm\) 2 & 242\\
\end{tabular}
\end{ruledtabular}
\end{table}

The state selectivity of ionization was verified using the QND state-detection scheme outlined in Section~\ref{subsec:QND}. In the present experiments, the state of seventy-two single N$_2^+$ ions was measured after ionization in the RF trap. Twenty-seven ions were detected in the rotational ground state implying a total state-preparation fidelity of $38\pm7~\%$. These include ions produced by threshold photoionization as well as by the non-state-selective processes quantified in Table~\ref{tab:Resonanse_proove}. 

As discussed in detail in References~\cite{zhang23a, blackburn20a}, RF and static electric fields inside the ion trap can lead to shifts of the ionization thresholds. If in such a scenario the ionization thresholds leading to excited rotational levels of the ion are shifted below the total photon energy provided by the ionization lasers, multiple ionic rotational levels can be produced compromising the fidelity for the preparation of the rotational ground state. Thus, the state-preparation fidelity critically depends on the instantaneous field strength during ionization. The amplitude of the oscillating RF fields in the trap is typically much larger than the strength of the static fields, thus the state-preparation fidelity is expected to primarily depend on the phase of the RF field at the time of ionization~\cite{blackburn20a}. 

To characterize the influence of the phase of the RF field on the state-preparation fidelity, the trigger of the ionization laser pulses was synchronized to the phase of the RF source of the trap. Simultaneously, RF amplitude and the potential on the endcap electrodes were reduced to $V_{RF}\approx237$~V and $V_{DC}=5$~V, respectively, to minimize further the influence of the fields on the ionization thresholds. The axial motional frequency for an N$_2^+$ ion corresponding to these parameters was $\omega_x = 148$ kHz at a Mathieu parameter~\cite{willitsch12a} $q=0.05$.

\begin{figure}[t]
\includegraphics[width=0.5\textwidth]{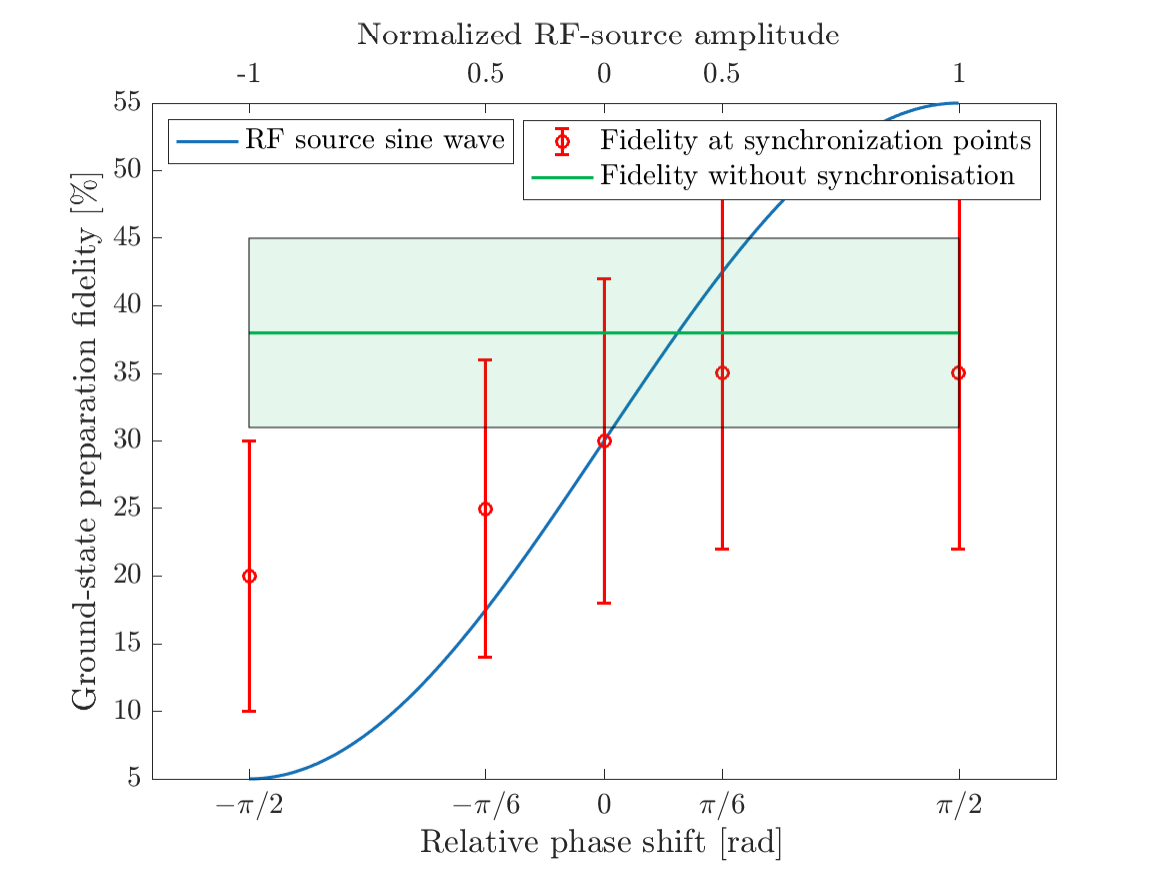}
\caption{Dependence of the state-preparation fidelity of photoionization on the relative phase of the ionization-laser trigger with respect to the RF source of the ion trap. The green trace indicates a value of 38\% rovibrational-ground-state preparation fidelity measured without synchronization to the RF source within its uncertainty limits (green shaded area). The red data points indicate the ground-state preparation fidelity at five different fixed phases of the RF cycle delivered by the source (blue line). The green shaded area and red error bars are Poissonian standard deviations. See text for details.}
\label{fig:REMPI_synchro}
\end{figure}

Due to the unknown phase shift between the RF source and the effective RF field inside the ion trap, the laser trigger was scanned with respect to the nominal phase of the RF source in order to identify the optimal working point. The maximum rovibrational-state-preparation fidelity is expected when the ionization laser pulse impinges on the molecular beam at a time around the zero crossing of the RF cycle. State-preparation fidelities at five different synchronization phases across an RF cycle were examined. At every phase, the state of twenty N$_2^+$ ions generated by photoionization nominally above the $N^+=0$ threshold was measured inside the ion trap. As can be seen in Figure~\ref{fig:REMPI_synchro}, at none of the five synchronization points an increase of the ground-state preparation fidelity above the value achieved without synchronization was obtained within the uncertainty limits. We thus conclude that the state-preparation fidelity is effectively independent of the RF phase under the conditions of the present experiment and the uncertainty limits of the measurement.

\section{\label{sec:discussion}Discussion}

\begin{figure*}[ht]
\includegraphics[width=1\textwidth]{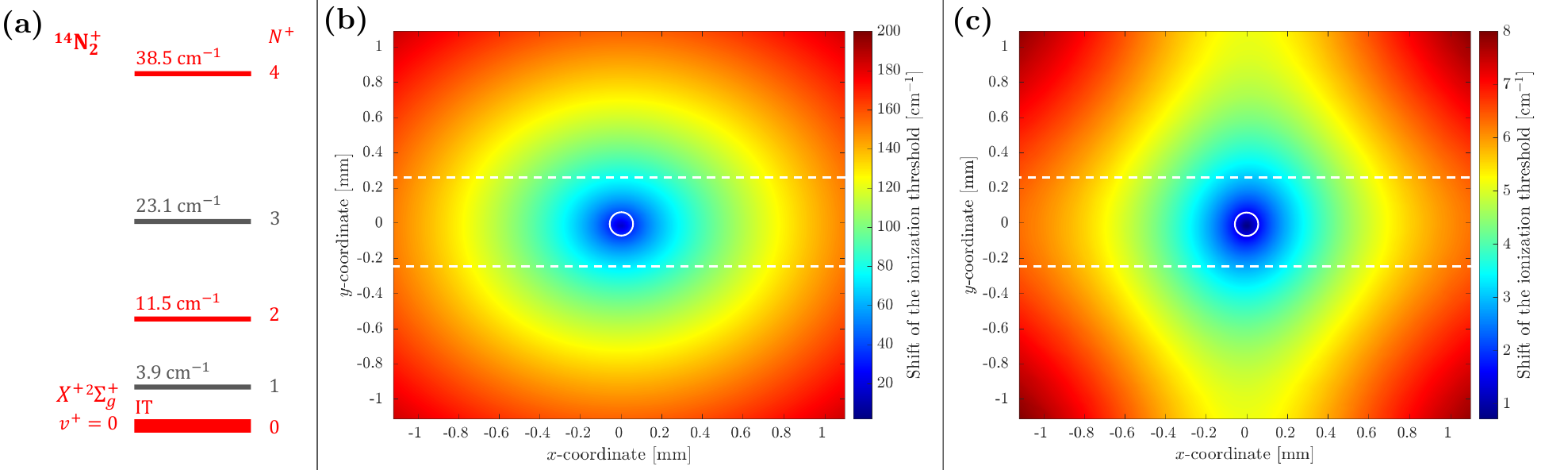}
\caption{a) Energy diagram of the rotational levels of N$_2^+$ in its $X^{+~2}\Sigma_g^+(v^+=0)$ vibronic ground state. Wavenumbers are referenced to the lowest ionization threshold (IT=125667.03~cm$^{-1}$~\cite{huber90a}). Levels which cannot be accessed in the present ionziation scheme for symmetry reasons are indicated in grey. b) and c) Calculated shifts of the ionization thresholds from (b) the RF field at maximum amplitude and (c) the static field produced by the endcap electrodes near the center of the trap (taken as the origin of the coordinate system). White circles and dashed white lines indicate the approximate boundaries of the laser focus and the molecular beam, respectively. See text for details.}
\label{fig:shifts}
\end{figure*}

The weak sensitivity of the ground-state preparation fidelity to the phase of the RF field observed in Figure~\ref{fig:REMPI_synchro} is striking and requires further analysis. To evaluate the shifts of the ionization thresholds of the rotational levels in the N$_2^+$ $X^{+2}\Sigma_g^+(v^+=0)$ ground state Figure~\ref{fig:shifts}a, the electric field at maximum RF amplitude and the static-field contribution produced by the trap endcaps were simulated using the COMSOL Multiphysics software~\cite{comsol6} assuming the trapping configuration employed experimentally during phase-synchronized ionization. The shifts of the ionization thresholds $\Delta$~[cm$^{-1}$] were calculated according to $\Delta=\alpha\sqrt{E}$, where a worst-case scenario of $\alpha=6.1$~cm$^{-1/2}$V$^{-1/2}$~\cite{merkt97a, chupka93a} was assumed and $E$~[V/cm] is the electric field experienced by the molecule. Figure~\ref{fig:shifts}b,c shows the spatial variation of the shifts of the ionization thresholds inside the trap in the plane along the propagation direction of the molecular beam ($X$-axis) and the vertical trap axis ($Y$-axis) caused by the RF and static electric fields, respectively. One can see that for the molecular beam with a diameter of $\approx500~\mu$m passing through the center of the trap and the ionization laser, focused to a spot size of $\approx150~\mu$m diameter in the center of the trap perpendicularly to the molecular beam, the shift on the edge of the ionization region reaches $\approx39~$cm$^{-1}$ and $\approx1.7~$cm$^{-1}$ from the RF and static field, respectively. Therefore, we conclude that the main contribution to the shift of the ionization thresholds in the present experiment is indeed produced by the RF field.

The effect of the synchronization of the zero crossing of the RF field with the timing of the ionization event has to be put into the context of the RF period of the trap compared to the duration of the ionization laser pulse. For our trap frequency of $\Omega_{RF}=2\pi\times16.4$~MHz, the half-period of the RF-field wave is 30.5~ns. Given that the temporal width at half maximum of the ionization laser pulse is $\tau\approx5$~ns, one obtains a spread in the phase of $\tau\times\Omega_{RF}=0.16\times 2\pi$. That means that even if the maximum of the laser pulse was perfectly synchronized with the RF zero crossings, ionization events occurring "on the edges of the pulse" still experienced about $25~\%$ of the full RF amplitude. Apparently, such events occurred sufficiently frequently in the experiments to mitigate the effect of the synchronization. This effect has been discussed theoretically in Reference~\cite{blackburn20a} and was confirmed experimentally here.

Generally, as shown here and also discussed elsewhere~\cite{zhang23a, blackburn20a} the fidelity of rotational-state preparation using threshold photoionization in traps strongly depends on the specific field configuration of the trapping environment. In large "soft" traps, state-preparation fidelities in threshold ionization exceeding 90\% have been demonstrated~\cite{tong10a, zhang23a}. However, in smaller traps with "stiff" trapping potentials which are required in experiments involving the cooling of the translational motion of the ions into the quantum regime~\cite{wolf16a, chou17a, sinhal20a, meir19a}, as in the present study, field effects become more important. 

In such cases, complete elimination of field effects could be obtained by temporarily quenching the RF potentials inside the trap in order to ionize the molecules in a completely field-free regime, as proposed in Reference~\cite{blackburn20a}. Alternatively, the ions could be generated outside the trap under field-free conditions and subsequently be inserted into the RF field. The efficiency of such approaches remains to be demonstrated. However, the state-preparation fidelities of about 40\% achieved here are readily high enough to serve as a starting point for an additional measurement-based state-selection step~\cite{najafian20a}. The state measurements on the single trapped ions using the scheme described in Section~\ref{subsec:QND} project the molecules into a specific quantum state which can be determined with near-unit ($>99$\%) fidelity as shown in Reference~\cite{sinhal20a}. Indeed, any ion measured to be in a specific state is \emph{known} to be in that state after the measurement and is thus available for subsequent state-selected experiments. Conversely, ions which are found not to be in the target state can be discarded and the experiment re-initialized until an ion in the appropriate state is detected. 

In this spirit, threshold photoionization represents a fitting approach to increase the initial state purity of the ions after which all ions not found to be in the desired state are eliminated by post-selection. This approach can increase the duty cycle of the experiment by orders of magnitude compared to studies which rely on ions prepared with thermal state populations at room temperature~\cite{wolf16a,chou17a} and thus crucially enhance the sensitivity of  experiments that require state-selected ions. 


\section{\label{sec:Summary}Summary}
In the present study, we evaluated strategies for the preparation of single molecular ions in well-defined rotational quantum states in an ion trap. We tested a $[2+1']$ REMPI scheme recently proposed by Gardner et al.~\cite{gardner19a} for the generation of trapped N$_2^+$ ions in their rotational ground state using a novel non-destructive state-detection scheme to characterize the fidelity of the state preparation on the single-ion level. Under the present experimental conditions, $38\pm7~\%$ of the ions produced were found in the rotational ground state limited by the influence of the inhomogeneous time-varying RF field in the ion trap and non-selective ionization processes. As the state of the ion is generally preserved in the measurements, ions found not to be in the target state can be discarded from the trap, leaving only state-selected ions for subsequent experiments. Thus, threshold photoionization combined with post-selection of the ions after QND state detection can serve as a highly efficient, widely applicable approach for experiments requiring state-selected molecular ions.

\section{\label{sec:Summary}Data availability}
The data that support the findings of this study are openly available on Zenodo at DOI 10.5281/zenodo.8273494.

\begin{acknowledgments}
The authors acknowledge support from the Swiss National Science Foundation (grant nr. 200021\_204123) and the University of Basel.

\end{acknowledgments}

\bibliography{REMPI_bibliography}

\end{document}